\documentclass{elsart}

\newcommand{\be}{\begin{equation}}
\newcommand{\ee}{  \end{equation}}
\newcommand{\ba}{\begin{eqnarray}}
\newcommand{\ea}{  \end{eqnarray}}

\usepackage{graphicx}
\usepackage{amsmath}
\usepackage{amssymb}

\journal{Annals of Physics}

\begin{document}

\begin{frontmatter}

\title{Interaction of Regular and Chaotic States}
\author{A. De Pace} and
\author{A. Molinari}
\address{Dipartimento di Fisica Teorica dell'Universit\`a di Torino and \\ 
  Istituto Nazionale di Fisica Nucleare, Sezione di Torino, \\ 
  via P.Giuria 1, I-10125 Torino, Italy}
\author{H. A. Weidenm\"uller}
\address{Max-Planck-Institut f\"ur Kernphysik, D-69029 Heidelberg, Germany}

\begin{abstract}
Modelling the chaotic states in terms of the Gaussian Orthogonal
Ensemble of random matrices (GOE), we investigate the interaction of
the GOE with regular bound states. The eigenvalues of the latter may
or may not be embedded in the GOE spectrum. We derive a generalized
form of the Pastur equation for the average Green's function. We use
that equation to study the average and the variance of the shift of
the regular states, their spreading width, and the deformation of the
GOE spectrum non--perturbatively. We compare our results with various
perturbative approaches.
\end{abstract}


\end{frontmatter}

\section{Introduction}

Predictions of the Gaussian Orthogonal Ensemble of Random Matrices
(GOE) offer the best first guess of spectral fluctuation properties of
a system about which nothing is known beyond the fact that it is
time--reversal invariant~\cite{Guh98}. Such a description is
completely adequate if the system is classically chaotic. Sometimes,
additional but incomplete dynamical information exists which calls for
an extension of the GOE. To make the need for such an extension
plausible, we mention several examples. (i) In spherical atomic
nuclei, the shell model is known to provide a proper dynamical
description of the low--lying states. Near neutron threshold, however,
the density of states in medium--mass and heavy nuclei is so high and
the levels are of such complexity that the shell--model approach
becomes unfeasible (even though it is probably still adequate in
principle). Moreover, there is evidence that the spectral fluctuation
properties of these states coincide with GOE predictions. The states
are accordingly modelled in terms of the GOE. The various realizations
of the GOE may then be viewed as corresponding to different choices of
the residual interaction of the shell model. The proper (i.e.,
realistic) choice is unknown. How do such different choices affect the
predictions of the shell model at lower excitation energies? The
question can be explored by studying an extension of the GOE
comprising in addition to the GOE Hamiltonian a set of discrete states
with energies below the continuous GOE spectrum. The states interact
with each other and with the GOE Hamiltonian. (ii) In
condensed--matter theory, a quantum dot carrying a single state may be
coupled to a finite reservoir of interacting electrons. The reservoir
may be modelled by a GOE Hamiltonian. The single state on the dot lies
within the GOE spectrum. How is the state affected by the interaction
with the GOE? That same question arises in finite Fermi systems when a
particular mode of excitation, which is not an eigenstate of the
Hamiltonian, occurs in a sea of states carrying the same quantum
numbers. Lacking a better description, those states are modelled in
terms of the GOE. In nuclear physics and in quantum chemistry, the
single state is referred to as a doorway state.

To address these and related questions, we study in the present paper
the spectral properties of a generic system consisting of a GOE
Hamiltonian interacting with a set of discrete states. The states may
or may not lie within the GOE spectrum. Related problems have been
studied by several authors. The doorway state mechanism has found wide
interest. Early reviews may be found in Refs.~\cite{Bo,Mah69}, see also
Ref.~\cite{Fes92}. The interaction between the higher--lying complex states and
the low--lying states of the shell model was, for instance, addressed in a
series of papers by Feshbach and collaborators~\cite{Fes96} and taken
up again in Refs.~\cite{Mol04}. The present work goes beyond these
papers in that we present a comprehensive treatment of the problem
within the framework of random--matrix theory.

The paper is organized as follows. The random--matrix model is defined
in Section~\ref{prob}. A straight perturbative approach which is useful
for purposes of orientation is described in Section~\ref{pert}, both
for non--overlapping and for overlapping spectra.
Section~\ref{sec:pastur} is the central theoretical piece of the
paper. We derive and analyze an extension of the Pastur equation for
the average Green's function of the system in the limit of infinite
matrix dimension of the GOE Hamiltonian. In Section~\ref{onedim} we
specialize that equation to the case of a single state interacting
with the GOE. We solve the generalized Pastur equation perturbatively.
In Section~\ref{sec:results} we present numerical results based upon
an exact solution of the Pastur equation. These are compared with the
various analytical approximations derived earlier.
Section~\ref{sec:concl} contains our conclusions. Some technical
details are deferred to the Appendix.

\section{Formulation of the Problem}
\label{prob}

We consider a decomposition of Hilbert space into two subspaces
defined by orthogonal projection operators $P = P^{\dag}$ and $Q =
Q^{\dag}$, with $P^2 = P$, $Q^2 = Q$, $PQ = 0$. The dimension of
$P$--space is $M$, that of $Q$--space is $N$. We shall assume that $N
\gg 1$ but (for reasons given below) will restrict ourselves to small
values of $M$. The dynamics in $P$--space is determined by the
Hamiltonian $H_0$. The motion in $Q$--space is assumed to be irregular
or stochastic and described in terms of random--matrix theory. The
Hamiltonian $H^{\mathrm{GOE}} = QH^{\mathrm{GOE}}Q$ is a member of the
ensemble of Gaussian orthogonal random matrices (GOE). The matrix
elements $H^{\mathrm{GOE}}_{\mu \nu}$ with $\mu, \nu = 1, \ldots, N$
are Gaussian--distributed random variables with zero mean value and
second moment \be \left\langle H^{\mathrm{GOE}}_{\mu \nu}
H^{\mathrm{GOE}}_{\rho \sigma} \right\rangle = \frac{\lambda^2}{N}
\bigg( \delta_{\mu \rho} \delta_{\nu \sigma} + \delta_{\mu \sigma}
\delta_{\nu \rho} \bigg) \ .
\label{1}
\ee
Here and in what follows, the average over the ensemble is denoted by
angular brackets, and $2 \lambda$ denotes the radius of Wigner's
semicircle, see Eq.~(\ref{P6}) below. The Hamiltonians in $P$--space
and in $Q$--space are coupled by an interaction $V = PVQ + QVP$. The
total Hamiltonian has the form
\begin{equation}
\label{eq:H}
  H = H_0 + V + H^{\mathrm{GOE}} \ .
\end{equation}
We study how the stochastic dynamics in $Q$--space affects the
dynamics in $P$--space.

\section{Perturbative Approach}
\label{pert}

It is instructive to use first a perturbative approach as this gives
some insight into the behavior of the system. We consider the case
$M = 1$. The single state in $P$--space carries the index $0$ while
the states in $Q$--space are labelled $\mu = 1, \ldots, N$. The matrix
representation of the Hamiltonian (\ref{eq:H}) reads
\begin{equation}
\label{eq:HP1}
  H = \left( \begin{array}{cc}
    E_0 & \widetilde{V}_{\nu} \\
    \widetilde{V}_{\mu} & H_{\mu\nu}^{\mathrm{GOE}}
    \end{array} \right)
\end{equation}
where $E_0 = (H_0)_{00}$, and where $\widetilde{V}_{0 \mu} =
\widetilde{V}_{\mu 0} = \widetilde{V}_{\mu}$ are the matrix elements
coupling the $P$--space to the $Q$--space. It is convenient to
diagonalize $H^{\mathrm{GOE}}$.  We call the eigenvalues $E_\mu$ and
the transformed coupling matrix elements $V_\mu$. The eigenvalues
$E_\mu$ obey Wigner--Dyson statistics, and the $V_\mu$'s are
uncorrelated Gaussian random variables with zero mean value and with a
common variance ${\mathcal V}^2$. Moreover, the $V_\mu$'s and the
$E_\mu$'s are uncorrelated. Eq.~(\ref{eq:HP1}) takes the form
\begin{equation}
  H = \left( \begin{array}{cc}
    E_0 & V_\nu \\
    V_\mu & E_\mu\delta_{\mu\nu}
    \end{array} \right).
\label{eq:HP2}
\end{equation}

\subsection{Non--overlapping Spectra}

By definition, the GOE spectrum is centered at $E = 0$. We assume that
the spectra of $H_0$ and of $H^{\mathrm{GOE}}$ do not overlap. Hence,
the distance $|E_0|$ of the $P$--space state from the centre of the
GOE spectrum must be larger than $2 \lambda$, the radius of the semicircle, see
Eq.~(\ref{P6}) below.
The secular equation for $H$ is easily found to read
\begin{equation}
\label{eq:secular}
  E_0 - \alpha = \sum_{\mu=1}^N \frac{V_\mu^2}{E_\mu-\alpha} \ .
\end{equation}
We assume that $({\mathcal V}^2)^{1/2}$ is small in comparison with
the difference between $E_0$ and the closest end point of the
semicircle. We accordingly write
\begin{equation}
  \alpha = E_0 + \delta \alpha
\end{equation}
and assume that we may solve Eq.~(\ref{eq:secular}) by expanding in
powers of $\delta \alpha$. This gives
\begin{equation}
  -\delta \alpha = \sum_{\mu=1}^N V_\mu^2 \sum_{n=0}^\infty
  \left(\frac{1}{E_\mu-E_0} \right)^{n+1} (\delta \alpha)^n
\end{equation}
which shows that $\delta \alpha$ is of order $\mathcal{V}^2$. Keeping
terms up to fourth order in $V_\mu$ we get
\begin{equation}
\label{eq:-y}
  -\delta \alpha \approx \sum_{\mu=1}^N \frac{V_\mu^2}{E_\mu-E_0}
  \left( 1 - \sum_{\nu=1}^N \frac{V_\nu^2}{(E_\nu-E_0)^2} \right) \ .
\end{equation}
We calculate both, the ensemble average $\left\langle \delta \alpha
\right\rangle$ and the variance $\left\langle( \delta \alpha -\langle
\delta \alpha \rangle)^2\right\rangle$ of $\delta \alpha$. Different
realizations of the GOE--Hamiltonian give rise to different values of
$\delta \alpha$. The ensemble average of $\delta \alpha$ yields the
mean shift of $E_0$ due to the interaction with the states in
$Q$--space, and the variance of $\delta \alpha$ is a measure of the
fluctuation of the position of the state in $P$--space due to
different realizations of the GOE Hamiltonian. The calculation of both
quantities uses $N \gg 1$ and is sketched in the Appendix. We recall
that for $|E| \leq 2 \lambda$,
\be
\rho(E) = \frac{N}{\pi \lambda} \sqrt{1 - (E/(2 \lambda))^2}
\label{P6}
\ee
is the average density of states of the GOE, and that the usual
definition of the spreading width $\Gamma^{\downarrow}$ for a state
mixed with the GOE and located at $E = 0$ is \cite{Fes92}
\be
\Gamma^{\downarrow} = 2 \pi {\mathcal V}^2 \rho(0) \ . 
\label{P0}
\ee
With this definition, we find
\be
  - \langle \delta \alpha \rangle = \frac{\Gamma^{\downarrow}}{2 \pi}
  I(x_0) \left[1-\frac{\Gamma^{\downarrow}}{4 \pi \lambda} \frac{\d
  I(x_0)}{\d x_0} \right] \ .
\label{P1}
\ee
Here $x_0 = E_0 / 2 \lambda$ obeys $|x_0| > 1$, and
\ba
\label{eq:I0}
  I(x_0) &=& \int_{-1}^1\d x \frac{\sqrt{1-x^2}}{x-x_0} \nonumber \\
  &=& 2\sqrt{x_0^2-1}\left[
    \mathrm{arctan}\left(\frac{1+x_0}{\sqrt{x_0^2-1}}\right) -
    \mathrm{arctan}\left(\frac{1-x_0}{\sqrt{x_0^2-1}}\right) \right]
    \theta(|x_0|-1) \nonumber \\ && - \pi x_0 \ .
\ea
The second term on the right--hand side of Eq.~(\ref{P1}) suggests
that the perturbation expansion proceeds in powers of
$\Gamma^{\downarrow} / \lambda$. For $|x_0| \gg 1$ we have $I(x_0)
\approx - \pi / (2 x_0)$ and
\be
   - \langle \delta \alpha \rangle \cong - \frac{\Gamma^{\downarrow}
   \lambda}{2 E_0} \ .
\label{P2}
\ee
The shift is always away from the center of the semicircle and of
order $\Gamma^{\downarrow}$.

For the variance of $\delta \alpha$ we obtain
\ba
\mathrm{var}(\delta \alpha) &=& \left\langle( \delta \alpha -\langle
\delta \alpha \rangle)^2\right\rangle \nonumber \\
&=& \frac{2 \pi}{N} \left( \frac{\Gamma^{\downarrow}}{2 \pi}
\right)^2 \bigg\{ \frac{\rm d}{{\rm d} x_0} I(x_0) -
\frac{\Gamma^{\downarrow}}{4 \pi \lambda} \left( \frac{\rm d}{{\rm d}
x_0} I(x_0) \right)^2 \nonumber \\
&& \qquad - \frac{\Gamma^{\downarrow}}{8 \pi \lambda} I(x_0)
\frac{{\rm d}^2}{{\rm d} x^2_0} I(x_0) \bigg\} \ .
\label{P3}
\ea
Eq.~(\ref{P3}) confirms the impression that the perturbation expansion
proceeds in powers of $\Gamma^{\downarrow} / \lambda$. Moreover we see
that $ (\mathrm{var}(\delta \alpha))^{1/2} / \left\langle \delta
\alpha \right\rangle$ is of order $N^{-1/2}$. This is due to the fact
that in Wick--contracting the $V_\mu$'s, we reduce the number of
independent summations over the $Q$--space states by one. For $|x_0|
\gg 1$ and to lowest order in $\Gamma^{\downarrow} / \lambda$, we find
\be
\sqrt{\mathrm{var} ( \delta \alpha )} \cong \frac{1}{\sqrt{N}}
\frac{\Gamma^{\downarrow} \lambda}{2 |E_0|} \ .
\label{P3a}
\ee
Eqs.~(\ref{P2}) and (\ref{P3a}) are in accord with the findings of
Ref.~\cite{Mol04}. 

The interaction between the levels in $P$--space and those in
$Q$--space induces an interaction amongst the levels in $P$--space. To
estimate that interaction we consider two degenerate $P$--space states
at energy $E_0$ and use a representation of the GOE Hamiltonian in the
form of Eqs.~(\ref{eq:HP1}) and (\ref{eq:HP2}). 
Perturbatively, the matrix element of the induced interaction between these two
states (labelled $a$ and $b$, respectively) is given by
\be
V^{\rm ind}_{a b} = \sum_\mu \frac{V_{a \mu} V_{\mu b}}{E_0 - E_\mu} \ . 
\label{P3b}
\ee
Using the same steps and notation as before, we find for the ensemble
average of $V^{\rm ind}_{a b}$ the approximate value $\langle
\tilde{V}_a | \tilde{V}_b \rangle / E_0$. With $\phi$ the angle between
the vectors $\tilde{V}_{a \mu}$ and $\tilde{V}_{b \mu}$ this is
approximately equal to $\cos(\phi) (\lambda/E_0)(\Gamma^\downarrow /
2)$. Comparing this expression with the average shift~(\ref{P2}) we
see that the averaged induced interaction matrix element is smaller by
the factor $\cos \phi$. Because of the complexity of the states in
$Q$--space, it is reasonable to expect that for $N \gg 1$ the two
vectors $\tilde{V}_a$ and $\tilde{V}_b$ are approximately orthogonal.
Then the induced interaction between two degenerate states in
$P$--space is very small.  If $\cos \phi = 0$ the variance of $V^{\rm
ind}_{a b}$ is small of order $1/N$. This case applies whenever we
assume that the two $P$--space states are coupled with equal strength
to the states in $Q$--space, see Sections~\ref{onedim} and
\ref{sec:results}. We conclude that it it reasonable to assume that
the induced interaction between the states in $P$--space is small in
comparison to the shift of each state. This is why we confine
ourselves in Section~\ref{sec:results} mainly to a one--dimensional
$P$--space.

\subsection{Overlapping Spectra}

We turn to the case where the spectra of $P$--space and $Q$--space
overlap, i.e., where $|E_0| < 2 \lambda$. Here the use of perturbation
theory may be somewhat doubtful but seems justified by the results. We
start from the perturbative solution~(\ref{eq:-y}) of the secular
equation~(\ref{eq:secular}). Taking the ensemble average, we change
the summation over the discrete energies $E_\mu$ into an integration
over a continuous variable $E^-$. The negative imaginary increment is
needed since the spectra overlap, and since the imaginary part of
$\delta \alpha$ is required to be negative. Formally, we obtain the
same expressions as in Eqs.~(\ref{P1}) and (\ref{P3}) except that
$I(x_0)$ must be replaced everywhere by
\be
I^-(x_0) = \int_{-1}^1 \ {\rm d} x \frac{\sqrt{1-x^2}}{x^- - x_0} \ .
\label{P4}
\ee
To lowest order in ${\mathcal V}^2$, this yields
\begin{equation}
\label{eq:yemb}
  -\langle \delta \alpha \rangle = - \frac{\Gamma^{\downarrow} E_0} {4
  \lambda} + i \frac{\Gamma^{\downarrow}}{2} \frac{\rho(x_0)}{\rho(0)}
  \ .
\end{equation}
The real part of the average shift vanishes for $E_0 = 0$ and
increases monotonically in magnitude as $E_0$ moves towards the end
points of the semicircle. It is negative (positive) for $E_0 < 0$
($E_0 > 0$), reflecting the effect of the level repulsion of the
$Q$--space states below and above $E_0$.

The interaction $V$ mixes the state in $P$--space with the states in
$Q$--space. Since the spectra overlap, that mixing is strong even for
small values of the interaction. The degree of mixing is measured by
the spreading width which has the expected value $2 \pi {\mathcal V}^2
\rho(E)$. For sufficiently small values of $\Gamma^{\downarrow} /
\lambda$, the probability of finding the $P$--state wave function
admixed to the true eigenstates of the system is described by a
Lorentzian with width $\Gamma^{\downarrow}$~\cite{Bo}. Obviously, the
spreading width phenomenon has no analogue in the case of non--overlapping
spectra. The quantity $\langle \delta \alpha \rangle$ gives the mean
values of shift and width both of which actually fluctuate about these
mean values. We have not calculated the fluctuations.

\section{Generalized Pastur Equation}
\label{sec:pastur}

In the present Section we use a non--perturbative approach to assess
the influence of the $Q$--space states onto the states in $P$--space.
We do so for $N \gg 1$. The approach makes use of a generalized form
of the Pastur equation.

\subsection{Average Green's Function}

The central element of our analysis is the retarded propagator $G(E)$
(or Green's function) of the system, averaged over the GOE. It is
defined as
\be
\left\langle G(E) \right\rangle = \left\langle \frac{1}{E^+ - H}
\right\rangle \ .
\label{green}
\ee
Here $E$ is the energy of the system and the plus indicates a positive
imaginary increment. From $\langle G(E) \rangle$, we find the average
level density (or spectral function) of the system as
\be
\rho = - (1/\pi) {\rm Im} \ {\rm Tr} \left\langle G(E) \right\rangle \ ,
\label{2}
\ee
where Tr stands for the trace.
The function $\langle G(E) \rangle$ obeys the generalized Pastur
equation
\begin{equation}
  \left\langle G(E) \right\rangle = G_0(E) + G_0(E) \left\langle
  H^{\mathrm{GOE}} \langle G(E) \rangle H^{\mathrm{GOE}} \right\rangle
  \langle G(E) \rangle \ .
\label{pastur}
\end{equation}
Here
\begin{equation}
  G_0(E) = \frac{1}{E^+ - H_0 - V}
\end{equation}
is the propagator of the system without any dynamics in $Q$--space.
Eq.~(\ref{pastur}) is obtained by expanding $\langle G(E) \rangle$ in
powers of $H^{\mathrm{GOE}}$, using Wick contraction, keeping only
leading terms in an asymptotic expansion in inverse powers of $N$
(these are the ``nested'' contributions), and resummation.
Alternatively, Eq.~(\ref{pastur}) can also be derived using
supersymmetry~\cite{Efe97,Guh98} and the saddle--point
approximation (the saddle--point equation coincides with the
generalized Pastur equation). We use Eq.~(\ref{1}) and the definition
\begin{equation}
\label{eq:sigma}
  \sigma = \frac{\lambda}{N} \mathrm{Tr} \left( Q \langle G(E) \rangle
  Q \right)
\end{equation}
to write Eq.~(\ref{pastur}) in the form
\be
  \left\langle G(E) \right\rangle = G_0(E) + \lambda \sigma G_0(E) Q
  \left\langle G(E) \right\rangle \ .
\label{Pastur}
\ee
To solve Eq.~(\ref{Pastur}), we project that equation onto $Q$-space
and obtain
\begin{equation}
\label{eq:QGQ}
  Q \left\langle G(E) \right\rangle Q = \frac{QG_0(E)Q}{1-\lambda
  \sigma QG_0(E)Q} \ .
\end{equation}
Taking the trace, we find for $\sigma$ the equation
\begin{equation}
  \label{sigma}
  \sigma = \frac{\lambda}{N} \mathrm{Tr}\left(\frac{QG_0(E)Q}
    {1-\lambda\sigma QG_0(E)Q} \right)
\end{equation}
which can be solved provided $QG_0(E)Q$ is known. Inserting the
solution into Eq.~(\ref{eq:QGQ}) yields the $Q$--space projection of
the ensemble--averaged propagator. The other projections of
$\langle G(E) \rangle$ are easily found, too. We get
\begin{subequations}
\label{G}
\begin{eqnarray}
  \label{eq:QGP} Q \left\langle G(E) \right\rangle P &=&
  \frac{1}{1-\lambda\sigma QG_0(E)Q} QG_0(E)P \ , \\ \label{eq:PGQ} P
  \left\langle G(E) \right\rangle Q &=& PG_0(E)Q (1+\lambda\sigma
  QG_0(E)Q) \ , \\ \label{eq:PGP} P \left\langle G(E) \right\rangle P
  &=& PG_0(E)P \nonumber \\ && + \lambda\sigma PG_0(E)Q \frac{1}{1 -
  \lambda\sigma QG_0(E)Q} QG_0(E)P \ .
\end{eqnarray}
\end{subequations}
The various projections of $G_0(E)$ which are needed in the
calculation, can be obtained using the expansion
\begin{equation} G_0(E) = \frac{1}{E^+ - H_0}
\sum_{n=0}^\infty \left( V\frac{1}{E^+ - H_0} \right)^n.
\end{equation}
This yields
\begin{subequations}
\label{G_0}
\begin{eqnarray}
  \label{eq:QG_0Q} QG_0(E)Q &=& \frac{1}{E^+ - QVP\frac{\displaystyle
  1}{\displaystyle E^+ - H_0} PVQ} \ , \\ \label{eq:QG_0P} QG_0(E)P
  &=& \frac{1}{E^+ - QVP\frac{\displaystyle 1}{\displaystyle E^+ -
  H_0} PVQ} QVP \frac{1}{E^+ - H_0} \ , \\ \label{eq:PG_0Q} PG_0(E)Q
  &=& \frac{1}{E^+ - H_0} PVQ \frac{1}{E^+ - PVQ\frac{\displaystyle
  1}{\displaystyle E^+} QVP} \ , \\ \label{sin} PG_0(E)P &=&
  \frac{1}{E^+ - PVQ\frac{\displaystyle 1}{\displaystyle E^+} QVP} \ .
\end{eqnarray}
\end{subequations}
These equations have an obvious physical interpretation.

\subsection{Equation for $\sigma$}

The Green's function $\langle G(E) \rangle$ is completely known if we
know $\sigma$, the solution of Eq.~(\ref{sigma}). To rewrite that
equation in a more explicit form we need to work out $QG_0(E)Q$.
Expanding $QG_0(E)Q$ in powers of $V$, rearranging the series and
resumming it, we obtain
\be
QG_0(E)Q = \frac{1}{E^+ - K} \ . 
\label{3}
\ee
Here $K$ is a matrix in $Q$--space given by
\be
K = Q V \frac{1}{E^+ - H_0} V Q \ .
\label{4}
\ee
The matrix $K$ is complex symmetric and can be diagonalized by an
energy--dependent complex orthogonal transformation. We denote the
complex and energy--dependent eigenvalues by $\kappa_j$. The imaginary
part of $K$ is negative semidefinite. Therefore, ${\rm Im} \ \kappa_j \leq
0$ for all $j$. The number of nonzero eigenvalues $\kappa_j$ is $\leq
M$. To see this, we use in $P$--space a basis in which $H_0$ is
diagonal and has eigenvalues $E_j$, $j = 1, \ldots, M$. Then,
\be
K_{\mu \nu} = \sum_j V_{\mu j} \frac{1}{E^+ - E_j} V_{j \nu} \
.
\label{5}
\ee
This shows that in $Q$--space and for fixed $E$, $K$ is a bilinear
form in the $M$ vectors $V_{\mu j}$ and, therefore, has rank $M$. We
also observe that the imaginary part of $K$ is nonzero only if $E$
coincides with one of the eigenvalues $E_j$. Therefore, the $M$
nonzero energy--dependent eigenvalues $\kappa_j$ are, in general, real
except for a set of discrete points. Using this diagonal form for $K$,
we rewrite Eq.~(\ref{sigma}) as
\be
\sigma = \frac{N - M}{N} \frac{\lambda}{E^+ - \lambda \sigma} +
\frac{1}{N} \sum_{j = 1}^M \frac{\lambda}{E^+ - \lambda \sigma -
\kappa_j} \ .
\label{6}
\ee

To discuss Eq.~(\ref{6}) we first consider the case where $P$--space
and $Q$--space are uncoupled. Then, $K = 0$ and $\kappa_j = 0$ for all
$j$, and Eq.~(\ref{6}) reduces to the well--known saddle--point
equation of the GOE which reads
\be
\sigma = \frac{\lambda}{E - \lambda \sigma} \ .
\label{7}
\ee
That equation yields
\begin{equation} \label{Q4}
  \sigma_0 = \frac{E}{2\lambda} \pm \mathrm{i}
  \sqrt{1-\left(\frac{E}{2\lambda} \right)^2}
\end{equation}
and, thus, the semicircle law~(\ref{P6}) for the normalized average
level density of the GOE. To compare with Eq.~(\ref{7}) we rewrite
Eq.~(\ref{6}) as
\be
\sigma = \frac{\lambda}{E^+ - \lambda \sigma} + \frac{1}{N} \sum_{j
= 1}^M \frac{\lambda \kappa_j}{(E^+ - \lambda \sigma) (E^+ - \lambda
\sigma - \kappa_j)} \ .
\label{8}
\ee
The saddle--point equation for the GOE is modified. The additional
terms reflect the properties of the Hamiltonian in $P$--space and its
coupling to $Q$--space. While Eq.~(\ref{7}) is a quadratic equation in
$\sigma$ and can be solved analytically, Eq.~(\ref{8}) is of order $M
+ 2$ and can only be solved numerically.  We use the following
strategy. We first deal with Eq.~(\ref{6}) in the case of a
one--dimensional $P$-space. Later we show that our main conclusions
are not affected as that dimension is increased.

\section{One--Dimensional $P$--Space}
\label{onedim}

\subsection{Basic Equations}

The dimension of Hilbert space is $N + 1$. As in Section~\ref{pert},
we denote the $P$--space component by the index zero while the indices
in $Q$--space run from $1$ to $N$. We write $(H_0)_{00} = E_0$ and
$V_{0 \mu} = V_{\mu 0} = V_\mu$. We rotate the system in $Q$-space,
exploiting the orthogonal invariance of the GOE, to have the vector
${\vec V}$, initially with components $V_\mu$, point in the direction
of the unit vector in the $N$--direction. We denote by $V$ the length
of that vector. We define
\begin{equation}
  \kappa_0 = \frac{V^2}{E^+ - E_0}
\end{equation}
and have from Eq.~(\ref{6})
\begin{equation}
\label{eq:sigma1}
  \sigma = \lambda \frac{N-1}{N} \frac{1}{E^+ - \lambda\sigma} +
    \frac{\lambda}{N} \frac{1}{E^+ - \lambda\sigma - \kappa_0} \ .
\end{equation}
For the projections of $G_0(E)$ we obtain
\begin{subequations}
\label{eq:Pastur1}
\begin{eqnarray}
\label{eq:QGQP1}
  (QG_0(E)Q)_{\mu\nu} &=& \delta_{\mu\nu} (1-\delta_{\mu N}) \frac{1}{E^+}
    + \delta_{\mu N} \delta_{\nu N} \frac{1}{E^+ - \kappa_0} \ , \\
  (QG_0(E)P)_{\mu1} &=& \delta_{\mu N} \frac{1}{E^+ - \kappa_0} V
    \frac{1}{E^+ - E_0} \ , \\
  (PG_0(E)Q)_{1\mu} &=& \delta_{\mu N} \frac{1}{E^+ - E_0} V
    \frac{1}{E^+ - \kappa_0} \ , \\
  (PG_0(E)P)_{11} &=& \frac{1}{E^+ - E_0 - \frac{\displaystyle V^2}
    {\displaystyle E^+}} \ .
\end{eqnarray}
\end{subequations}

Eq.~(\ref{eq:sigma1}) is the equation we explore numerically in the
next Section. We are mainly interested in the average level density as
given by Eq.~(\ref{2}). To this end we need aside from $\sigma$ (which
according to Eq.~(\ref{eq:sigma}) determines ${\rm Tr} Q \langle G(E)
\rangle$) also $\tau = \lambda P \langle G(E) \rangle P$. From
Eqs.~(\ref{G}) and (\ref{eq:Pastur1}), we find
\begin{equation} \label{tau}
  \tau = \lambda P \left\langle G(E) \right\rangle P = \frac{\lambda
  (E - \lambda \sigma)}{(E^+ - E_0)(E^+ - \lambda \sigma) - V^2} \ .
\end{equation}

\subsection{Perturbative solution of the Pastur equation}

To gain insight into the nature of the solutions, it is instructive to
solve the generalized Pastur equation for $\sigma$ perturbatively. We
write Eq.~(\ref{eq:sigma1}) in the form
\be
\label{eq:sigma1bis}
  \sigma = \frac{\lambda}{E - \lambda\sigma} \left[ 1 + \frac{1}{N}
    \frac{V^2}{(E-E_0)(E-\lambda\sigma)-V^2} \right] \ .
\ee
According to Eq.~(\ref{2}), the energies for which the cubic
equation~(\ref{eq:sigma1bis}) possesses a pair of complex conjugate
solutions $\sigma$ (or, equivalently, a single real solution) define
the spectrum of $H$.

\subsubsection{Nonoverlapping Spectra}

When $E_0$ lies outside the semicircle, we expect that the spectrum of
$H$ consists of two disconnected pieces: One piece should more or less
coincide with the range $-2 \lambda \leq E \leq 2 \lambda$ of the GOE
spectrum. We focus attention on the other piece which should cover a
small energy interval close to the point $E_0$. We assume $E_0 < - 2
\lambda$ and $|E_0| / \lambda \gg 1$.

The first factor on the right--hand side of Eq.~(\ref{eq:sigma1bis})
taken by itself would give rise to the unperturbed solution~(\ref{Q4})
which, for $E_0 < - 2 \lambda$ and $|E_0| / \lambda \gg 1$, is
approximately given by $\lambda / E_0$. We insert this value into the
first factor on the right--hand side of Eq.~(\ref{eq:sigma1bis}) and
solve the resulting quadratic equation in $\sigma$. We put $E = E_0$
everywhere except for the term $(E - E_0)$. This yields
\be
\sigma \cong \frac{1}{2 \lambda} \left( E_0 - \frac{V^2}{E -
E_0} \right) \pm \sqrt{\frac{1}{4 \lambda^2} \left( E_0  -
\frac{V^2}{E - E_0} \right)^2 - \frac{V^2}{N E_0 (E - E_0)}} \ .
\label{Q6}
\ee
The end points of the spectrum are those values of $E$ where the
argument of the square root vanishes. This yields
\be
E = E_0 + \frac{V^2}{E_0} \left( 1 \pm \frac{2 \lambda}{\sqrt{N} E_0}
\right) \ . 
\label{Q7}
\ee
To compare this with the result of Section~\ref{pert}, we recall the
definition~(\ref{P0}), the fact that $\rho(0) = N / (\pi \lambda)$,
and the fact that ${\mathcal V}^2$ is the average value of the
$V^2_\mu$'s while $V^2$ is their sum. Thus, $N {\mathcal V}^2 = V^2$.
We see that the center of the interval defined in Eq.~(\ref{Q7})
coincides with the shift~(\ref{P2}) while the length of the spectrum
is smaller by the factor $2 \lambda / |E_0|$ than the perturbative
result~(\ref{P3a}).

As in Section~\ref{pert} we briefly address the interaction induced
between two states in $P$--space labelled $a$ and $b$ by their
interaction with the states in $Q$--space. Again we assume that the
two states are degenerate with common energy $E_0$. A little algebra
shows that the relevant two--dimensional matrix is ${\bf M}_{\alpha
\beta} = \langle \tilde{V}_\alpha | \tilde{V}_\beta \rangle$ where
$\alpha$ and $\beta$ run from $a$ to $b$ and where we have used the
notation of Eq.~(\ref{eq:HP1}) (our statement agrees with the
arguments in Section~\ref{pert}). We denote the eigenvalues of ${\bf
M}_{\alpha \beta}$ by $m_a$ and $m_b$. For simplicity and lack of
detailed knowledge it is often assumed that the coupling of both
states to $Q$--space is the same so that $m_a = m_b$. In this case,
the Pastur equation takes the same form as Eq.~(\ref{eq:sigma1bis}),
with $V^2 = m_a$ but with $1/N$ replaced by $2/N$. Making that same
substitution in Eq.~(\ref{Q7}) we see that the induced level repulsion
is of order $1/\sqrt{N}$. This is in keeping with the result of
Section~\ref{pert} if we note that $m_a = m_b$ implies $\langle
\tilde{V}_a | \tilde{V}_b \rangle = 0$. When that inner product
differs from zero, $m_a$ and $m_b$ must necessarily differ. Expanding
the terms in the Pastur equation in powers of the difference $m_a -
m_b$, it is straightforward to show that the leading non--vanishing
contributions are of order $(m_a - m_b)^2$. That shows that
significant level repulsion sets in only slowly as $|{\bf M}_{a b}|$
increases from zero. We take this result as a further justification
for studying in Section~\ref{sec:results} mainly a one--dimensional
$P$--space.

\subsubsection{Overlapping Spectra}

When $E_0$ lies inside the semicircle, we expect the spectrum to
consist of a single stretch of the energy axis but with a strong
enhancement of the average level density for $E$ near $E_0$. We start
from the solution~(\ref{Q4}) of the unperturbed Pastur
equation~(\ref{7}) and write
\begin{equation}
  \sigma = \sigma_0 + \delta\sigma \ .
\end{equation}
Inserting this into Eq.~(\ref{eq:sigma1bis}) leads to
\begin{equation}
\label{eq:dsigma}
  \delta\sigma = \frac{\lambda}{N} \frac{V^2}{(E-2
  \lambda\sigma_0-\lambda \,\delta\sigma)[(E - E_0)(E -
  \lambda\sigma_0 - \lambda\,\delta\sigma)-V^2]} \ .
\end{equation}
This expression suggests that $\delta \sigma$ is of order $1/N$.
Hence, we neglect $\delta \sigma$ on the right--hand side. This
approximation does not take into account the shift of the end points
of the GOE spectrum due to the presence of the state in $P$--space. An
improved approximation is introduced below. We find
\be
\delta \sigma \cong \frac{\lambda}{N} \frac{V^2}{(E - 2
  \lambda \sigma_0 )[(E - E_0)(E - \lambda \sigma_0) - V^2]} \ .
\label{Q1}
\ee
Likewise we get for $\tau$ from Eq.~(\ref{tau})
\begin{equation}
  \tau \cong
  \frac{\lambda(E-\lambda\sigma_0)}{(E-E_0)(E-\lambda\sigma_0)-V^2} \
  .
\end{equation}
It is convenient to introduce dimensionless variables. We define
\begin{equation}
\label{eq:dimless}
  x = \frac{E}{2 \lambda}, \quad x_0=\frac{E_0}{2 \lambda}, \quad 
    \gamma = \frac{V^2}{\lambda^2} \ .
\end{equation}
As usual we consider the retarded Green's function and, therefore,
choose the negative sign on the right--hand side of Eq.~(\ref{Q4}).
We obtain
\begin{equation}
  \delta \sigma = - \mathrm{i} \frac{\tilde{\gamma}}{4 N} \frac{x -
  \mathrm{i} \sqrt{1 - x^2}}{\sqrt{1 - x^2}[x - x^{\mathrm{shift}}_0
  + (\mathrm{i} / 2) \tilde{\gamma} \sqrt{1-x^2}]}
\end{equation}
and
\begin{equation}
  \tau = \frac{1}{2 - \gamma} \ \frac{1}{x - x_0^{\mathrm{shift}} +
  (\mathrm{i}/2) \tilde\gamma \sqrt{1-x^2}} \ .
\end{equation}
Both $\delta\sigma$ and $\tau$ display a Breit--Wigner resonance at
the shifted energy
\begin{equation}
  x_0^{\mathrm{shift}} = \frac{2 x_0}{2 - \gamma} = x_0 +
  \frac{\tilde\gamma}{2} x_0 \ ,
\end{equation}
with a width $\tilde\gamma \sqrt{1-x^2}$ where
\begin{equation}
  \tilde\gamma = \frac{2 \gamma}{2-\gamma} \ .
\end{equation}
We note that $\tilde\gamma$ is a nonlinear function of $\gamma$. To
the best of our knowledge, that non--linearity has never been taken
into account in applications of the doorway--state model to data. The
singularity of $\tilde{\gamma}$ at $\gamma = 2$ (which corresponds to
$\Gamma^{\downarrow} = 4 \lambda$) occurs when the spreading width
equals the diameter of the semicircle and is, thus, far beyond
reasonable applications of the model. To lowest order in $\gamma$,
both shift and width agree with the perturbative result of
Section~\ref{pert}.

The contribution to the average level density stemming from the
$P$--space state is given by the sum of the imaginary parts of
$N\delta\sigma$ and of $\tau$. Upon integration over all energies
these should add up to unity. This is not the case, however, because
our approximation fails at the end points of the semicircle. Thus, the
present approximation is useful only if the distance of $E_0$ from the
closest end point is large compared to $V^2 / \lambda$.

We improve on this approximation by writing the generalized Pastur
equation~(\ref{eq:sigma1bis}) in the form
\be
\label{eq:sigma2bis}
  \sigma (E - \lambda\sigma) = \lambda \left[ 1 + \frac{1}{N}
    \frac{V^2}{(E-E_0)(E-\lambda\sigma)-V^2} \right] \ .
\ee
We consider this as a quadratic equation in $\sigma$ with known
right--hand side. This yields
\be
  \sigma = \frac{E}{2 \lambda} - i \sqrt{1 - \bigg( \frac{E}{2
  \lambda} \bigg)^2 + \frac{1}{N} \frac{V^2}{(E - E_0)(E -
  \lambda\sigma) - V^2}} \ .
\label{Q2}
\ee
An approximate solution is obtained by assuming that $N \gg 1$ and
substituting for $\sigma$ on the right--hand side the unperturbed
solution~(\ref{Q4}). Then
\be
\sigma \cong \frac{E}{2 \lambda} - i \sqrt{1 - \bigg( \frac{E}{2
  \lambda} \bigg)^2 + \frac{1}{N} \frac{V^2}{(E - E_0)(E / 2 +
  \mathrm{i} \lambda \sqrt{1 - (E / (2 \lambda))^2}) - V^2}} \ .
\label{Q3}
\ee
For $E_0 \approx 2 \lambda$ near the end point of the semicircle and
$E \geq E_0 \gg V^2 / \lambda$, the last term under the square root
gives a positive contribution: Because of level repulsion, the end
point of the spectrum is pushed away from the center. The argument
applies correspondingly when $E_0 \approx - 2 \lambda$. The resonance
contribution can be discussed along similar lines as before.

\section{ Numerical Results}
\label{sec:results}

The results of our numerical calculations lead to a deeper
understanding of the theory developed in the previous Sections. We
solve the generalized Pastur equation exactly. We use dimensionless
variables defined as
\be
y = \frac{E}{\lambda} \ , \ y_0 = \frac{E_0}{\lambda} \ , \ \gamma =
\frac{V^2}{\lambda^2} \ .
\label{Q5}
\ee We recall that the end points of the semicircle are located at
$\pm 2$. The definition~(\ref{Q5}) implies that $\gamma =
\Gamma^{\downarrow} / (2 \lambda)$. Therefore, physically reasonable
values of $\gamma$ obey $\gamma \leq 1 / 2$ or so, and we have
restricted ourselves to that range.

\subsection{Non--overlapping Spectra}

\begin{figure}[t]
\begin{center}
\includegraphics[clip,height=0.63\textheight]{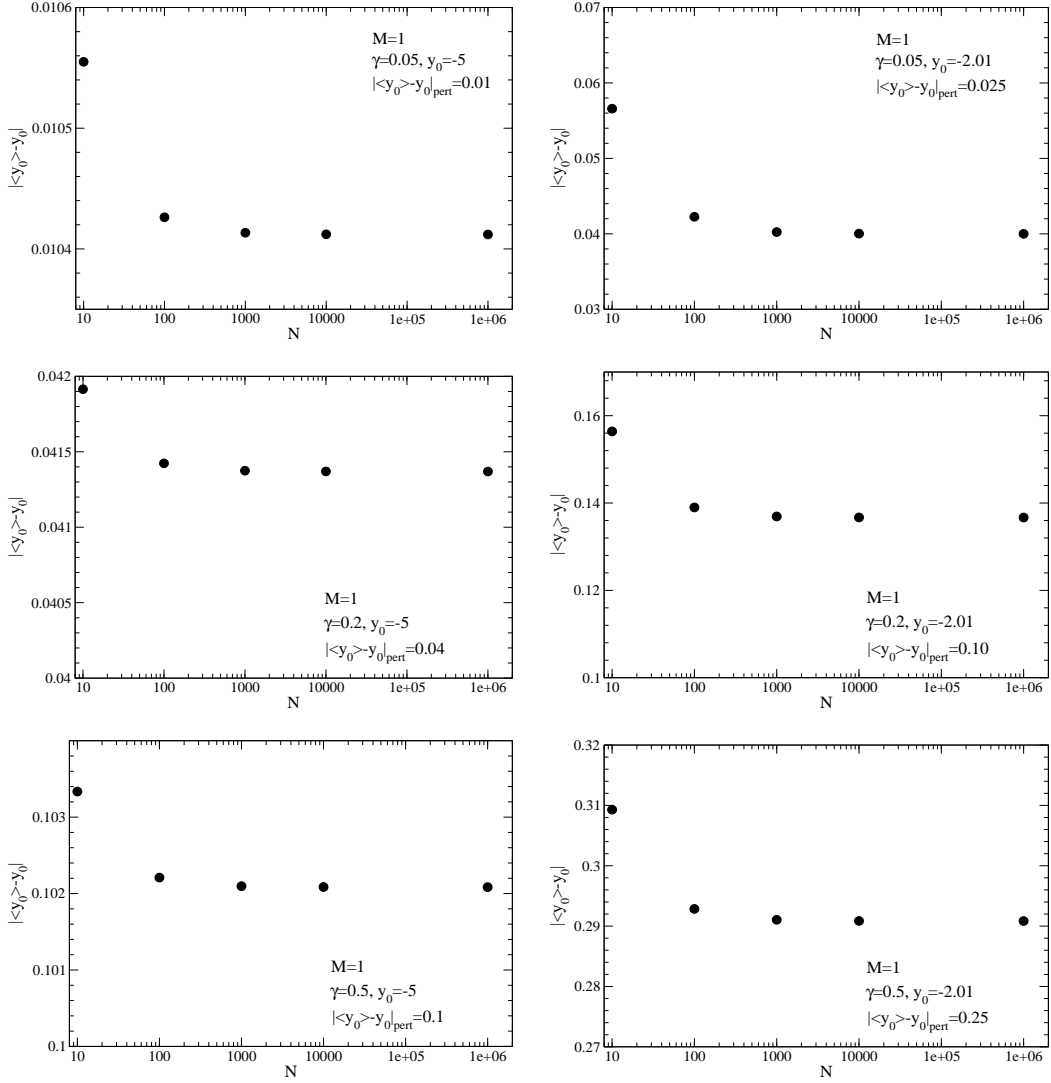}
\caption{\label{fig:P1_y_y0-2-5} The shift of the single $P$--space
state (M = 1) due to the coupling with $N$ $Q$--space states versus
$N$ for several values of the coupling strength $\gamma$ and for two
positions $y_0$ of the state.}
\end{center}
\end{figure}

We first consider the case of a one--dimensional $P$--space hosting
the system's ground state. The numerical calculation yields the end
points of that part of the spectrum which is due to the presence of
the $P$--space state. From here we calculate the center of the
spectrum and its length. We call the difference between the
unperturbed position and the center of the spectrum the shift and the
length of the spectrum the fluctuation and compare both values with
the perturbative estimates of Sections~\ref{pert} and \ref{onedim}.
We must keep in mind, of course, that the present definitions of shift
and fluctuation differ from the ones used in Section~\ref{pert}.

In Fig.~\ref{fig:P1_y_y0-2-5} the downward shift of the $P$--space
state due to the coupling with the $N$ chaotic $Q$--space states is
displayed versus $N$ for several values of the coupling strength
$\gamma$. For the sake of comparison, the $N$--independent
perturbative result $|\langle y_0 \rangle - y_0|_{\rm pert}$ is also
given in each panel. In all cases, the downward shift increases
monotonically with the number $N$ of states in $Q$--space and reaches
saturation at about $N = 10^4$. In the sequel, ``shift'' is used for
that asymptotic value.

\begin{figure}[t]
\begin{center}
\includegraphics[clip,height=6cm]{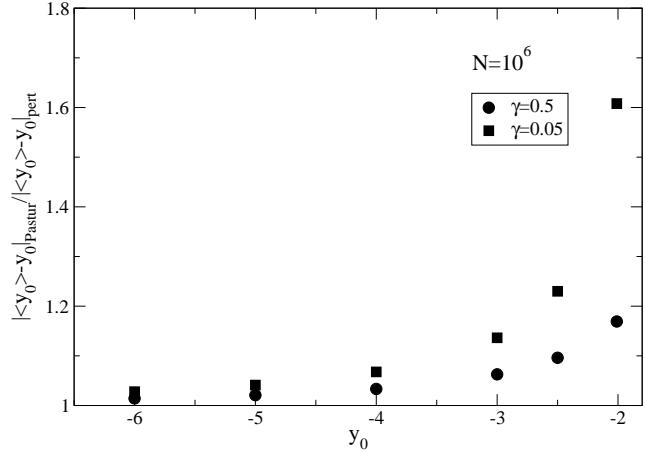}
\caption{\label{P1_ry_y0-2to6.ps} Ratios of exact and perturbative
results for the shift for $N \gg 1$ versus distance from the centre of
the semicircle.}
\end{center}
\end{figure}

In the Figure we display two cases:
\begin{itemize}
\item[a)] The $P$--space state lies far from the $Q$--space spectrum.
  The downward shift increases monotonically and almost exactly
  linearly with $\gamma$. Indeed, in going from $\gamma = 0.05$ to
  $\gamma = 0.5$ the value grows by an order of magnitude. The
  predictions of perturbation theory are in order and, notably, turn
  out to be always {\em smaller in magnitude} than the ones stemming
  from the generalized Pastur equation albeit by a small amount
  (roughly by about $4\div6\%$ for $N$ not too large).
\item[b)] The $P$--space state lies just below the semicircle. Again
  the shift increases monotonically with $\gamma$ but now perturbation
  theory badly underestimates the shift, especially for small values
  of $\gamma$.
\end{itemize}

\begin{figure}[t]
\begin{center}
\includegraphics[clip,height=0.63\textheight]{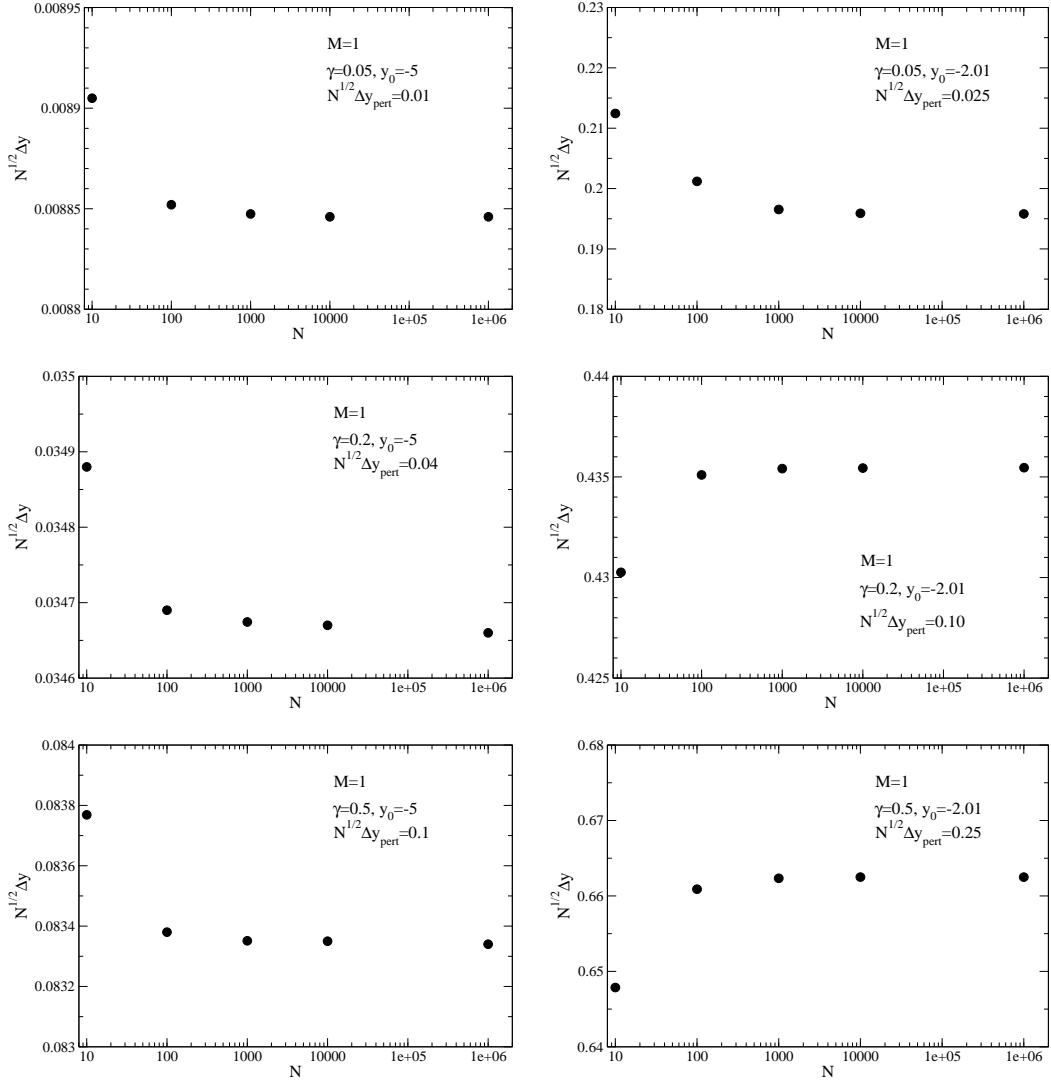}
\caption{\label{fig:P1_dy_y0-2-5} The fluctuation $\Delta y$ (square
root of the variance) of the position of the single $P$--space state
versus $N$ for several values of the coupling strength $\gamma$ and for
two positions $y_0$ of the state. We have scaled $\Delta y$ with
$N^{1/2}$.}
\end{center}
\end{figure}

A summary of our results is given in Fig.~\ref{P1_ry_y0-2to6.ps}. For
two values of $\gamma$, we show the ratio of the exact result over
the perturbative result for the shift versus distance from the centre
of the semicircle. We recall that in our units, the radius of the
semicircle is two. The Figure indicates the goodness of perturbative
results. For reasonable values of $\gamma \leq 1/2$, lowest--order
perturbation theory furnishes reliable results if the distance of the
$P$--state from the edge of the semicircle is larger than the radius
of the semicircle.

\begin{figure}[t]
\begin{center}
\includegraphics[clip,height=7.3cm]{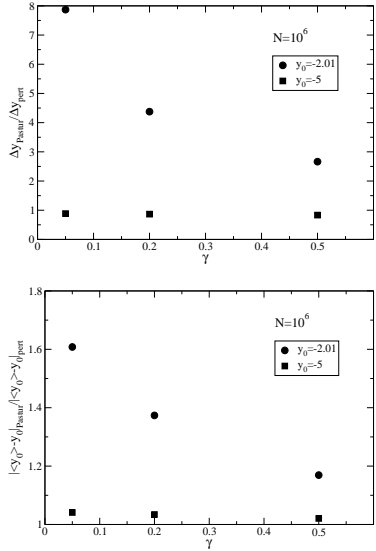}
\caption{\label{fig:P1_r_y0-2-5} Ratios of exact and perturbative
results for shift and fluctuation.}
\end{center}
\end{figure}

The fluctuations of the ground--state energy are displayed in
Fig.~\ref{fig:P1_dy_y0-2-5}. The perturbative result is also given in
each panel. In all cases, the fluctuations reach saturation at about
$N = 10^4$. In the sequel, ``fluctuation'' is used for that asymptotic
value.
\begin{itemize}
\item[a)] When the $P$--space state is far from the semicircle we find
  again that the fluctuations grow monotonically with $\gamma$, and
  that perturbation theory is reliable but, notably, yields results
  {\em larger} than the ones stemming from the Pastur equation (by
  about $10\div15\%$ when $\gamma$ is small, up to about $18\%$ when
  $\gamma$ is large; these estimates depend, of course, upon the definition
  of $\Delta y$).
\item[b)] Perturbation theory fails when the $P$--space state lies
   close to the semicircle yielding values much smaller than those
   inferred from the Pastur equation.
\end{itemize}

Our results are summarized in Fig.~\ref{fig:P1_r_y0-2-5} where we show
the ratios of the exact values over the perturbative results for the
fluctuations (upper panel) and for the shifts (lower panel) versus the
coupling strength $\gamma$. We observe that the discrepancy between
perturbative and exact results is decreasing with increasing $\gamma$.
We have no explanation for this observation.

\begin{figure}[t]
\begin{center}
\includegraphics[clip,height=7.5cm]{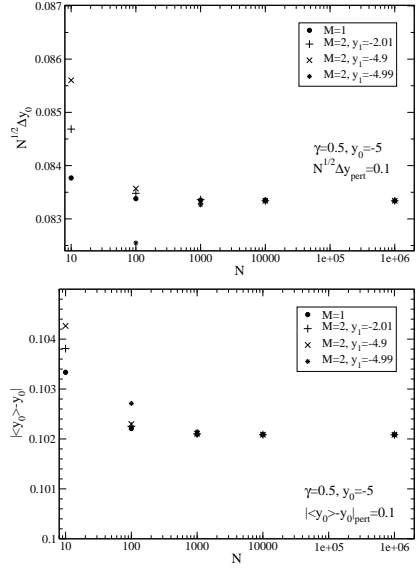}
\caption{\label{fig:P1P2_ydy_y0-5_gamma0.50} Influence of the induced
level repulsion between two states in $P$--space on shift and
fluctuation versus $N$.}
\end{center}
\end{figure}

To assess the validity of our results for realistic cases, we must
address the case of $P$--spaces of dimension $> 1$. We do so for a
two--dimensional $P$--space. We assume that the two $P$--space states
are not coupled to each other, and that both have the same average
coupling to the $Q$--space states. The results for this case are
displayed in Fig.~\ref{fig:P1P2_ydy_y0-5_gamma0.50}. We show versus
$N$ the average shift and the fluctuation of the lowest $P$--space
state located originally at $y_0 = -5$ for four cases: (i) $P$--space
is one--dimensional (M = 1), (ii) $P$--space is two--dimensional and
the second state is originally located at $y_1 = - 2.01$, (iii)
$P$--space is two--dimensional and the second state is originally
located at $y_1 = - 4.9$, (iv) $P$--space is two--dimensional and the
second state is originally located at $y_1 = - 4.99$, i.e., in the
immediate vicinity of the first state.

We observe that the downward shift of the lower of the two $P$--space
states increases as the distance to the higher $P$--space state is
reduced. In fact, the shift now arises from the combined action of the
coupling with the $Q$--space and the induced coupling to the other
state in $P$-space. However, this effect is visible only for $N \le
100$. For large values of $N$, the induced level repulsion between the
two states in $P$--space is negligible. This is in accord with the
results of Sections~\ref{pert} and \ref{onedim}. The fluctuations also
increase as the two states of the $P$-space come closer to each other
but again this effect is visible only for $N \le 100$.

\subsection{Overlapping Spectra}

\begin{figure}[p]
\begin{center}
\includegraphics[clip,height=0.90\textheight]{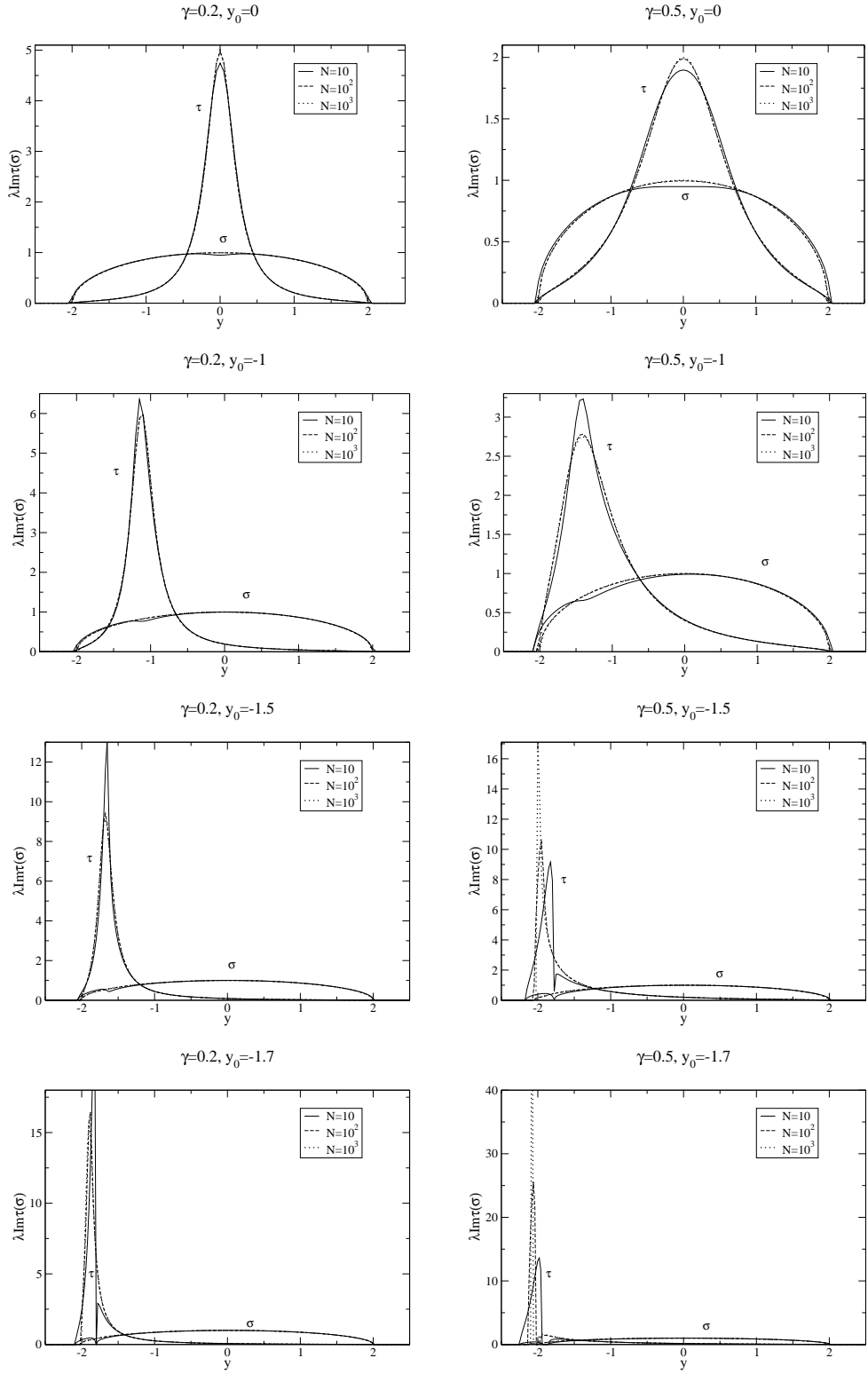}
\caption{\label{fig:PintoQa} The imaginary parts of $\lambda \tau$ and
of $\lambda \sigma$ for two different coupling strengths $\gamma$, and
for several positions $y_0$ of the $P$--space state.}
\end{center}
\end{figure}

\begin{figure}[p]
\begin{center}
\includegraphics[clip,height=0.90\textheight]{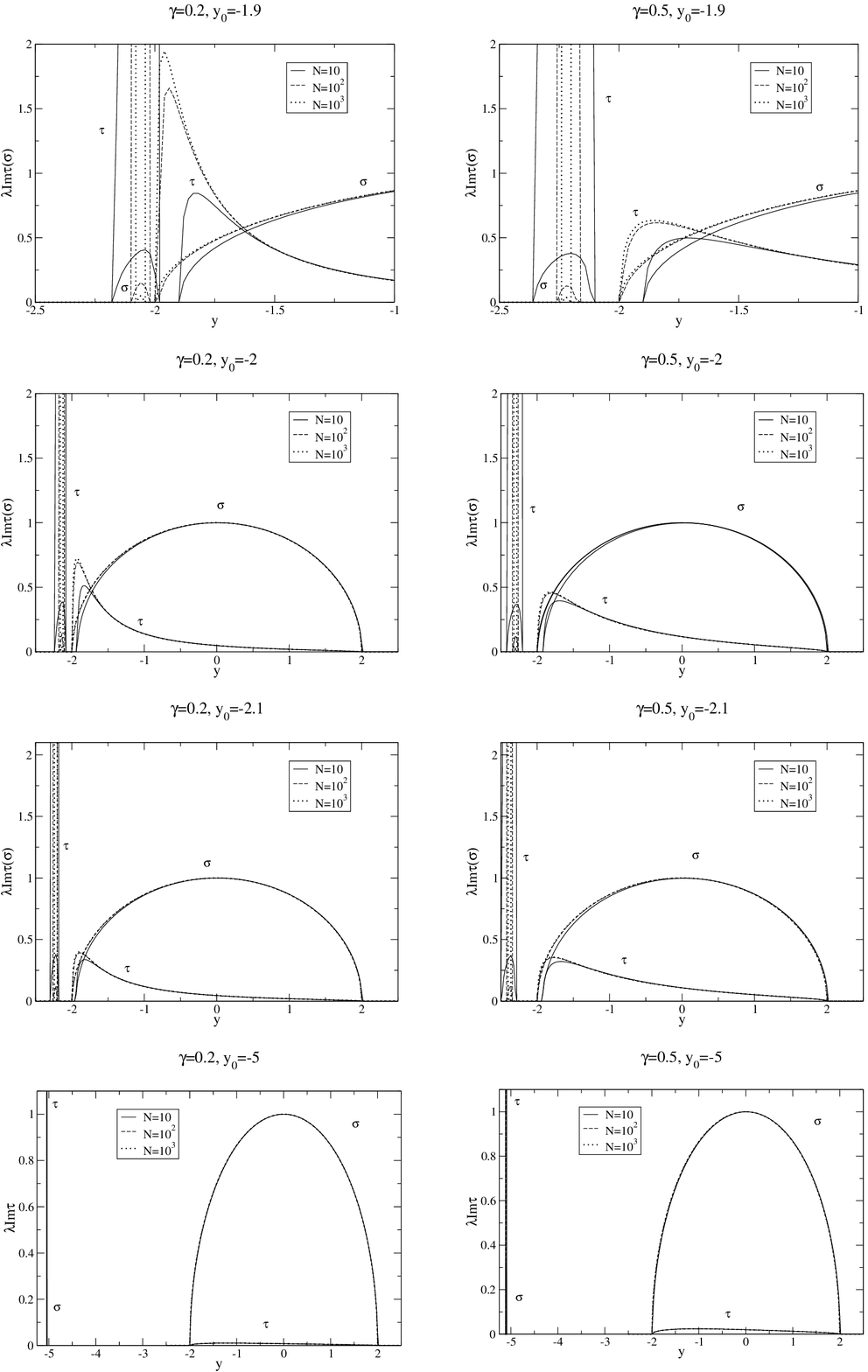}
\caption{\label{fig:PintoQb} Same as Fig.~\protect\ref{fig:PintoQa} but for 
$y_0$ close to the end point of the semicircle. The four panels at the bottom 
show the values of $\lambda \tau$ and of $\lambda \sigma$ for the case of
non--overlapping spectra.}
\end{center}
\end{figure}

We consider a one--dimensional $P$--space with $E_0$ embedded into the
GOE spectrum. We display the imaginary parts of the spectral functions
$\lambda \tau$ and $\lambda \sigma$ as defined in Eqs.~(\ref{tau}) and
(\ref{eq:sigma}). In addition we also show these two functions for the
case where $E_0$ lies far outside the semicircle.

Figs.~\ref{fig:PintoQa} and \ref{fig:PintoQb} show the imaginary parts
for two values of the coupling strength $\gamma$ and for several
values of the initial position $y_0$ of the $P$--space state. In each
panel we display the results for several choices of the dimension $N$ of
$Q$--space. We see that in the weak--coupling regime ($\gamma =
0.2$) and for $y_0 = 0$, the $P$--space spectral function $\tau$ is
well represented by a Lorentzian. As $y_0$ moves towards the end point
of the GOE spectrum, the Lorentzian becomes increasingly distorted.
With the Lorentzian normalized to $\pi$ the product of height and width
(computed at half its maximum) should be 2. For the figure shown that product 
is larger than 2 (the more so the more $y_0$ approaches the center of the
semicircle). As a consequence of level repulsion, the imaginary part of
$\sigma$ is not confined to the interior of the original GOE semicircle. For
$y_0$ near the center of the semicircle, the dilatation of the semicircle
gets smaller, however, with increasing $N$. The situation changes when
$y_0$ is close to the end point $-2$.

\begin{figure}[p]
\begin{center}
\includegraphics[clip,height=9.35cm]{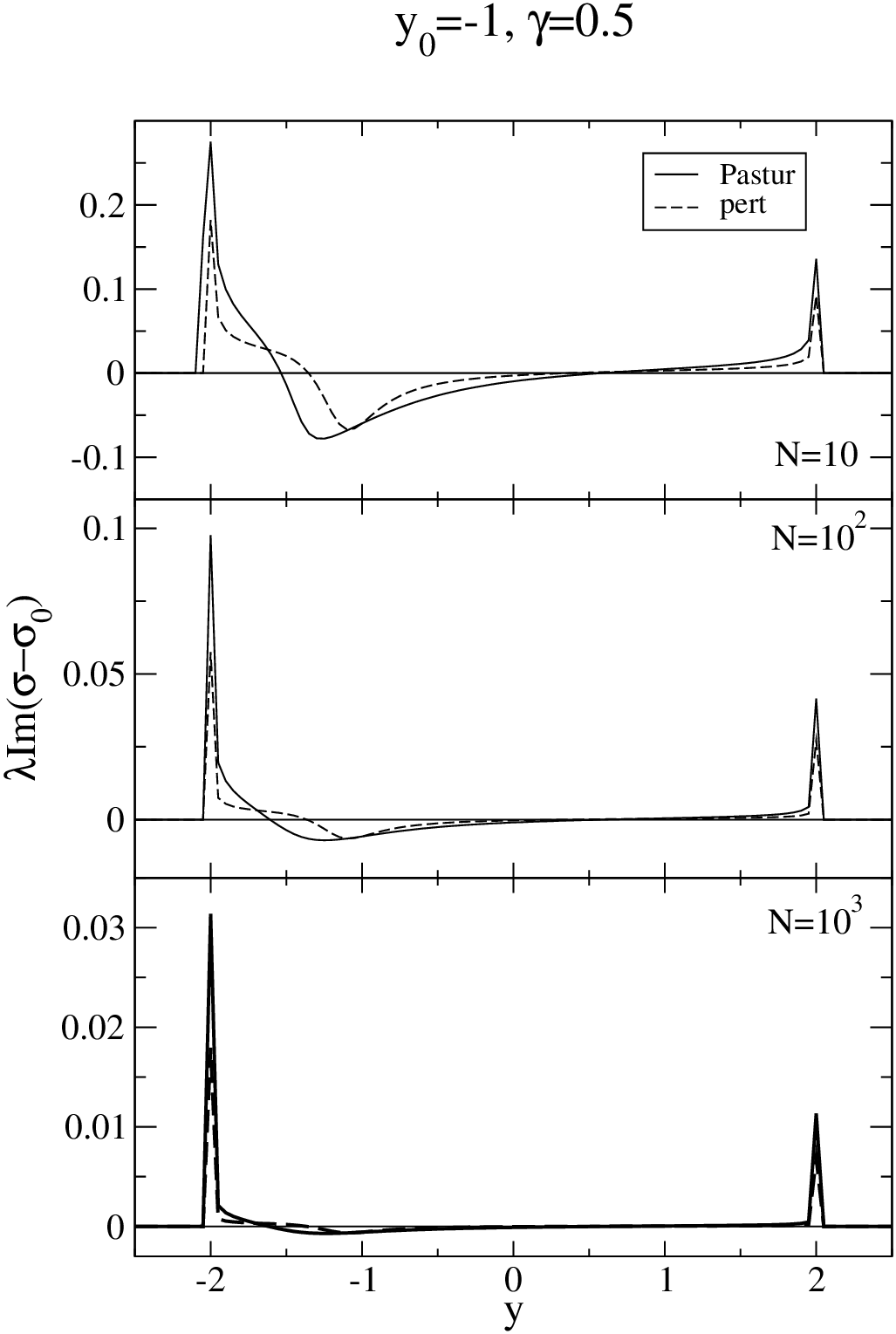}
\caption{\label{P1_sigmap_y-1.0_gamma0.50.ps} $\lambda {\rm Im}
  (\sigma-\sigma_0)$ as obtained from the exact solution of the Pastur equation
  (solid line) and from the perturbative solution (\ref{Q3}) (dashed line) for 
  $y_0=-1$, $\gamma=0.5$ and several values for $N$. }
\bigskip
\includegraphics[clip,height=9.35cm]{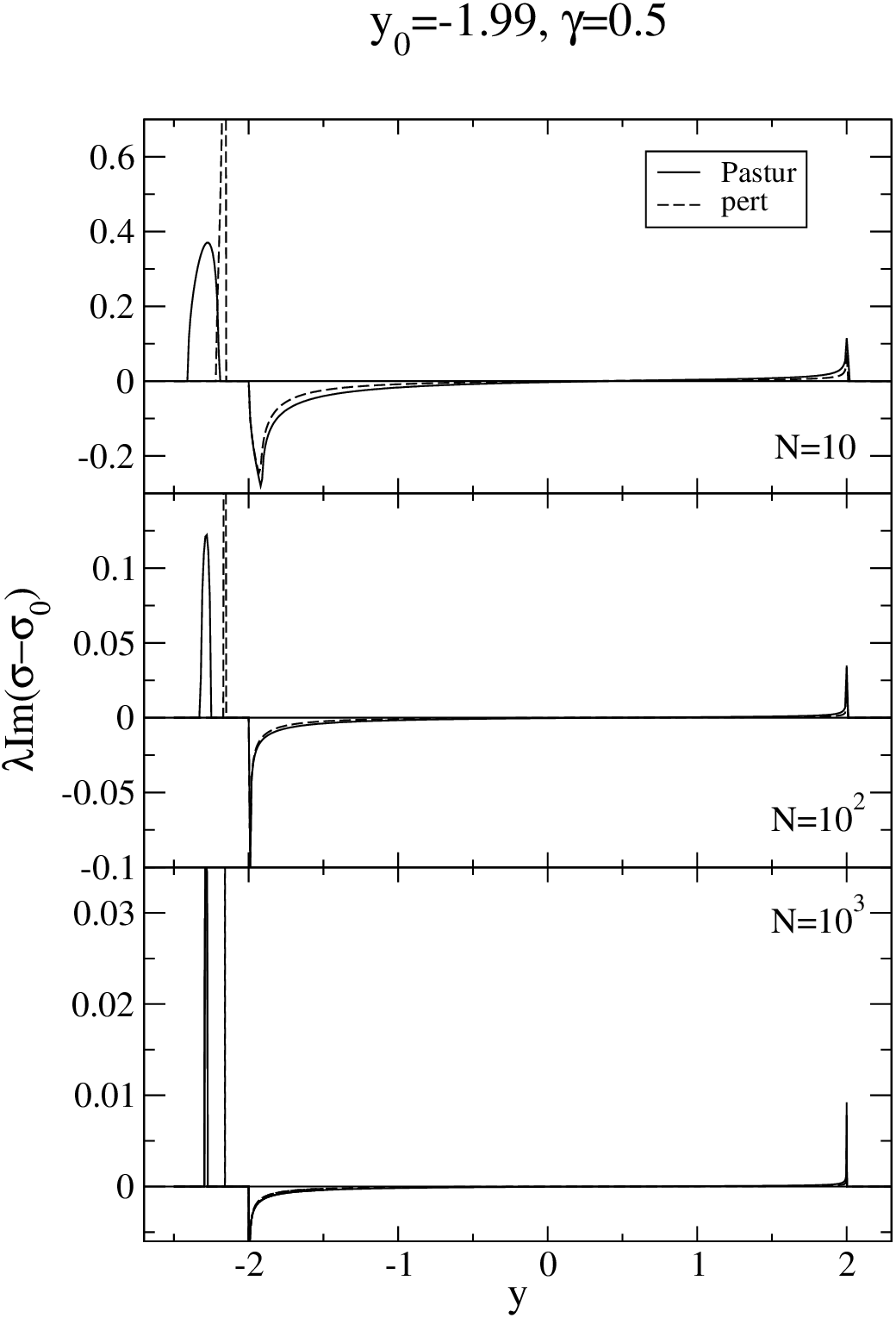}
\caption{\label{P1_sigmap_y-1.99_gamma0.50.ps} Same as
  Fig.~\protect\ref{P1_sigmap_y-1.0_gamma0.50.ps}, but for $y_0=-1.99$. }
\end{center}
\end{figure}

In the strong--coupling regime ($\gamma = 0.5$) and for $y_0$ near the
center of the semicircle, the imaginary part of $\tau$ resembles a
Gaussian more than a Lorentzian. Remarkably, the peak height is still given by
$1/\gamma$, irrespective of the strength of $\gamma$. Again, the function
becomes strongly distorted as $y_0$ approaches the end point. As a consequence
of level repulsion, the total spectrum may even develop two branches although
the $P$--space state originally lies within the semicircle.

It is of interest to compare the perturbative solution~(\ref{Q3}) of
the Pastur equation with the exact one. For two values of $y_0$
(position of the $P$--state) and several values of $N$, this is done
in Figs.~\ref{P1_sigmap_y-1.0_gamma0.50.ps} and
\ref{P1_sigmap_y-1.99_gamma0.50.ps}. We note that for $N \gg 1$, the
perturbative solution becomes amazingly accurate, even in the strong coupling
limit. However, when the $P$--space state is very close to the edge of the
semicircle ($y_0=-1.99$), perturbation theory appears to be unable to
reproduce the position of the peak outside the semicircle.

We have used the exact results partly displayed in Figs.~\ref{fig:PintoQa} and
\ref{fig:PintoQb} to check whether the imaginary parts of $N \sigma/\pi$ and 
$\tau/\pi$ add up to $N + 1$ as they should. 
We found this indeed to be the case to within about 0.1 percent. This
is the expected numerical accuracy of our results. 
Remarkably, this outcome is valid no matter what the value of $N$: We thus
conclude that Pastur equation, while missing contributions of the order of
$1/N$, yet preserves the normalization of the density of states.

\section{Summary and Conclusions}
\label{sec:concl}

We have given a comprehensive treatment of the interaction between
regular and chaotic states. We have modelled the latter in terms of
the GOE, the Gaussian Orthogonal Ensemble of Random Matrices. Using
the limit of large matrix dimension for the GOE, we have derived a
generalization of the Pastur equation for the ensemble average of the
Green's function of the system. That non--perturbative equation
furnishes exact results via numerical calculations. Using that
approach we have shown that the coupling induced between two regular
states by their interaction with the chaotic states is very small and,
in fact, negligible. This is true provided both states are coupled
equally strongly to the chaotic states. For that reason, we have
focused attention throughout most of the paper on a single regular
state interacting with a large number of chaotic states.

To gain an analytical understanding of the behavior of the system we
use approximate results. To this end we employ two types of
perturbation theory: Canonical perturbation theory for the secular
equation for the eigenvalues of the Hamiltonian, and a perturbative
treatment of the generalized Pastur equation.

We have considered two cases: The regular state lies either outside or
inside the GOE spectrum. In the first case, we calculate the mean
value and the variance of the shift, or another measure of its spread,
as mean values over the ensemble. The two perturbative approaches
yield essentially identical results and agree with the exact solution
if the distance between the energy of the regular state and the end
point of the GOE spectrum is sufficiently large. For realistic
coupling strengths that means the distance is similar to the radius of
the semicircle. Due to the interaction with the chaotic states, the
spectrum of a set of regular levels will get quenched. The quenching
is not uniform but grows with decreasing distance from the semicircle.

The interaction not only alters the position (and the wave function)
of the regular state but also affects the shape of the GOE spectrum.
The ensuing deformation is particularly relevant when the regular
state lies within the GOE spectrum and close to one of its end points.
The state itself gets shifted, and acquires a spreading width.
Canonical perturbation theory, based on a power series expansion in the
strength of the coupling between regular and chaotic states, 
yields estimates for these quantities.
The perturbative solution to the Pastur equation does not rely on the
smallness of that strength and displays a non--linear dependence of the 
spreading width on the strength. We believe this to be an interesting effect
which does not seem to have been taken into account previously and which
deserves further study. The shape of the spectrum due to the regular
state is Lorentzian for small couplings and gradually changes into a
Gaussian as the coupling is increased and becomes comparable to the
radius of the semicircle. The shape change of the chaotic spectrum is
not accessible to canonical perturbation theory but can be estimated
using the modified perturbation theory for the generalized Pastur
equation. Comparing the results of the latter with the exact ones, we
find that the perturbative solution is excellent for $N \gg 1$. 
When the energy of the regular state is close to the edge of the GOE
spectrum, the regular state is pushed outside the GOE spectrum as a 
consequence of level repulsion for sufficiently strong coupling.
In that case the spectrum develops two branches. The perturbative approach is
unable to account quantitatively for the part of the spectrum outside the
semicircle. 

In conclusion: We have derived and solved exactly the generalized Pastur
equation for a complex system with strong interactions. We have shown that
perturbation theory, either canonical or on the Pastur equation, is valid.
We may view our dynamical system as a generic model for a complex, strongly
interacting many--body system. In the light of such a view, our results are
remarkable. They convey the huge simplification of the many--body problem
induced by the concept of ensemble averaging.

\section*{Appendix}

To calculate the shift of the unperturbed energy $E_0$ we have to
evaluate the ensemble averages of the first and of the second term on
the right--hand side of Eq.~(\ref{eq:-y}). For the first term we
obtain
\begin{equation}
  \left\langle \sum_{\mu=1}^N \frac{V_\mu^2}{E_\mu-E_0} \right\rangle = 
    {\mathcal V}^2 \left\langle \sum_{\mu=1}^N \frac{1}{E_\mu-E_0}
    \right\rangle = {\mathcal V}^2 \int \d E \frac{1}{E-E_0}
    \sum_{\mu=1}^N \langle\delta(E-E_\mu)\rangle \ .
\end{equation}
The expression $\sum_{\mu=1}^N \langle\delta(E-E_\mu)\rangle$ is the
average level density of the GOE and given in Eq.~(\ref{P6}). With the
dimensionless variables $x=E/2\lambda$ and $x_0=E_0/2\lambda$ we get
\begin{equation}
  \left\langle \sum_{\mu=1}^N \frac{V_\mu^2}{E_\mu-E_0} \right\rangle = 
    \frac{{\mathcal V}^2 N}{\pi\lambda} I(x_0)
\end{equation}
where $I(x_0)$ is defined in Eq.~(\ref{eq:I0}). As for the second term
on the right--hand side of Eq.~(\ref{eq:-y}), we keep only the leading
contributions in an expansion in inverse powers of $N$. Thus, in the
identity $\langle V_\mu^2 V_\nu^2 \rangle = ({\mathcal V}^2)^2 (1 +
2\delta_{\mu\nu})$ the contribution from the Kronecker delta is
negligible. Averaging over the $V_\mu$'s yields
\begin{equation}
  \left\langle \sum_{\mu=1}^N \frac{V_\mu^2}{E_\mu-E_0} \sum_{\nu=1}^N
    \frac{V_\nu^2}{(E_\nu-E_0)^2} \right\rangle = ({\mathcal V}^2)^2
    \left\langle \sum_{\mu=1}^N \frac{1}{E_\mu-E_0} \sum_{\nu=1}^N
    \frac{1}{(E_\nu-E_0)^2} \right\rangle.
\end{equation}
The two--level correlation function decreases as a Bessel function
over distances large with respect to the level spacing $d$, but small
with respect to $\lambda\approx Nd$~\cite{Efe97} and can be neglected
in a calculation keeping only the leading order in $1/N$. Thus we get
\begin{equation}
  \left\langle \sum_{\mu=1}^N \frac{V_\mu^2}{E_\mu-E_0} \sum_{\nu=1}^N
    \frac{V_\nu^2}{(E_\nu-E_0)^2} \right\rangle = 
    \frac{({\mathcal V}^2)^2 N^2}{2\pi^2\lambda^3} I(x_0)
    \frac{\d I(x_0)}{\d x_0} \ .
\end{equation}
For the shift this gives Eq.~(\ref{P1}). The variance of $\delta
\alpha$ is obtained similarly.

\end{document}